%Paper: hep-th/9412082
%From: kibler@lyolav.in2p3.fr
%Date: Fri, 9 Dec 94 12:34:10 +0100

\input tex_inputs:maclyon
\magnification\magstep1
\baselineskip = 0.5 true cm
% \parskip=0.5 true cm

  \def\sa{\vskip 0.30 true cm}
  \def\sb{\vskip 0.60 true cm}
  \def\sc{\vskip 0.15 true cm}
  \def\sd{\vskip 0.50 true cm}
  \def\demi{ { {\lower 3 pt\hbox{$\scriptstyle 1$}} \over
               {\raise 3 pt\hbox{$\scriptstyle 2$}} } }

%  \pageno = 1
   \nopagenumbers
  \vsize = 25.7   true cm
  \hsize = 15.6   true cm
  \voffset = -1 true cm

\font\msim=msym10
\def\gr{\hbox{\msim R}}

\def\grn{\hbox{\msim N}}

\def\grc{\hbox{\msim C}}

\rightline{LYCEN 9439}
\rightline{October 1994}

\sc
\sb
\sb
\vskip 1 true cm

\centerline {{\bf SOME ASPECTS OF $q$- AND $qp$-BOSON
CALCULUS}{\footnote{$^*$} {\sevenrm Contribution to the Symposium
``Symmetries in Science VIII''
held at the Cloister Mehrerau (Bregenz, Austria, 7-12
August 1994). To appear in the
Proceedings of
the Symposium,
ed.~B.~Gruber
(Plenum Press, New York, 1995).}}}

\vskip 0.4 true cm

\sa
\sb

\vskip 0.5 true cm

\centerline
{M.R.~Kibler,
 R.M.~Asherova,\footnote{$^1$} {\sevenrm
 Permanent address~: Physics and Power Engineering
Institute, Obninsk, Kaluga region, Russia.}
and Yu.F.~Smirnov\footnote{$^2$} {\sevenrm
Present address~: Instituto de F\'\i sica,
Universidad Nacional Autonoma de M\'exico,
M\'exico D.F. 001, M\'exico.
On leave of absence from~:
% Permanent address~:
Institute of Nuclear Physics,
Moscow State University, 119899 Moscow, Russia.}}

\sa

\centerline {Institut de Physique Nucl\'eaire de Lyon}
\centerline {IN2P3-CNRS et Universit\'e Claude Bernard}
\centerline {43 Boulevard du 11 Novembre 1918}
\centerline {F-69622 Villeurbanne Cedex}
\centerline {France}

\sa

\sa
\sa
\sb
\sb

\centerline {\bf Abstract}

\sa
\sa

A set of compatible formulas for the Clebsch-Gordan coefficients of the
quantum algebra $U_{q}({\rm su}_2)$ is given in this paper. These formulas
are $q$-deformations of known formulas, as for instance: Wigner, van
der Waerden, and Racah formulas. They serve as starting points
for deriving various realizations of the unit
tensor of $U_{q}({\rm su}_2)$ in terms of $q$-boson operators. The
passage from the one-parameter quantum algebra $U_{q }({\rm su}_2)$
          to the two-parameter quantum algebra $U_{qp}({\rm  u}_2)$ is
discussed at the level of Clebsch-Gordan coefficients.

\baselineskip = 0.7 true cm

  \sa
% \sa
% \sa
% \sa
% \sb

\vfill\eject

\baselineskip = 0.5 true cm

% \vglue 5 true cm
  \vglue 4 true cm

\noindent {\bf SOME ASPECTS OF $q$- AND $qp$-BOSON CALCULUS}

\vskip 0.4 true cm

\sa
\sb

\vskip 0.5 true cm

\leftskip = 1.6 true cm

{M.R.~Kibler,
 R.M.~Asherova,\footnote{$^1$} {\sevenrm
Permanent address~: Physics and Power Engineering
Institute, Obninsk, Kaluga region, Russia.}
and Yu.F.~Smirnov\footnote{$^2$} {\sevenrm
Present address~: Instituto de F\'\i sica,
Universidad Nacional Autonoma de M\'exico,
M\'exico D.F. 001, M\'exico.
On leave of absence from~:
% Permanent address~:
Institute of Nuclear Physics,
Moscow State University, 119899 Moscow, Russia.}}

\sa

{Institut de Physique Nucl\'eaire de Lyon}

{IN2P3-CNRS et Universit\'e Claude Bernard}

{43 Boulevard du 11 Novembre 1918}

{F-69622 Villeurbanne Cedex}

{France}

\leftskip = 0 true cm

\sa
\sb
\sa
\sa
\sb
\sb

\baselineskip = 0.5 true cm

\noindent {\bf 1. PRELIMINARIES}

\sb

The aim of the present paper is to continue the program of extending
in the framework of $q$-deformations the main results of the work in
ref.~1 on the SU$_2$ unit tensor or Wigner operator (the matrix elements
of which are coupling coefficients or $3 - jm$ symbols).
A first part of this program
was published in the proceedings of {\it Symmetries in Science VI}
(see ref.~2)
where the $q$-deformed Schwinger algebra was defined and
where an algorithm,
based on the method of complementary $q$-deformed algebras,
was given for obtaining three- and four-term recursion relations for
the Clebsch-Gordan coefficients (CGc's) of
$U_q({\rm su}_{2  })$ and
$U_q({\rm su}_{1,1})$.  The algorithm was fully exploited in ref.~3
where the complementary of three quantum algebras in a
$q$-deformation of the symplectic Lie algebra sp($8,\gr$) was used for
producing 32 recursion relations.

This paper is organized as follows.
In section 2, we derive 12 explicit forms for the CGc's of
$U_q({\rm su}_{2  })$. They are $q$-deformations of the most usual
formulas displayed in the literature. [In the following, we shall use the
terminology {\it X form} which means that the corresponding formula can be
identified with the one originally derived by the author(s) {\it X}
({\it X}
$=$      Wigner,$^4$
van der Waerden,$^5$
          Racah,$^6$
       Majumdar,$^7$ etc.)
in the limiting case where $q=1$.]
{}From each of the 12 {\it X} forms,
it is possible to derive, as explained
in section 3, a $q$-boson realization
of the $U_q({\rm su}_{2  })$ unit tensor.
Finally,  section  4  deals  with  the
two-parameter Hopf algebra $U_{qp}({\rm u}_2)$
and it is sketched there how to transcribe to
    $U_{qp}({\rm  u}_2)$ the results obtained in section 2
for $U_{q }({\rm su}_2)$.

Some words about the notation are in order.
In section 4, we shall use the notation
$$
[[x]]_{qp} : = { {q^x - p^x} \over {q-p} }
\eqno (1)
$$
while in sections 2 and 3 we shall use the abbreviation $[x]$ to denote
$$
[x] \equiv [x]_q : = [[x]]_{qq^{-1}} = { {q^x - q^{-x}} \over {q-q^{-1}} }
\eqno (2)
$$
where $x$ may stand for an operator or a (real)
number. The other notations to be used concern
the $q$- and $qp$-deformed factorials
$$
\eqalign{
[n]! \ {\rm or} \ [n]_q! & := [1] [2] \cdots [n] \equiv [1]_q [2]_q \cdots
[n]_q
\qquad
     \quad [0]! \ {\rm or} \ [0]_q! := 1 \cr
[[n]]_{qp}! & := [[1]]_{qp} \,
                [[2]]_{qp} \, \cdots \,
                [[n]]_{qp}
\quad          \qquad [[0]]_{qp}! := 1
}
\eqno (3)
$$
where $n$ is a positive integer.

We shall take the commutation relations of the quantum algebra
 $U_q({\rm su}_{2  })$ in the usual (Kulish-Reshetikhin-Drinfeld-Jimbo) form,
viz.,
 $$
[J_3,J_- ] \ = \ - J_- \quad      \qquad
[J_3,J_+ ] \ = \ + J_+ \quad      \qquad
[J_+,J_- ] \ = \ [2J_3]_{q}
\eqno (4)
 $$
%     (4) = ex (1)
The co-product of the Hopf algebra
 $U_q({\rm su}_{2  })$ corresponds to eq.~(41) below with $p=q^{-1}$.
The extension of eq.~(4) to the quantum algebra $U_{qp}({\rm u}_2)$
is given in section 4.

For Hermitean conjugation requirements
[more precisely, to insure that $(J_+)^{\dagger} =
                                  J_-$],
the values of the parameters
$q$ and $p$ must be restricted to the following domains: either
(i)  $q \in \gr $ and
     $p \in \gr $ or
(ii)            $q \in \grc$ and
                $p \in \grc$ with $p=q^{\ast}$
(the ${\ast}$ indicates complex conjugation). In the special case where
$p=q^{-1}$, we may take either
(i)  $q \in \gr $ or
(ii) $q = {\rm e}^{{\rm i} \beta}$ with
$0 \le \beta < 2 \pi$. Therefore, in all cases the product $qp$ is real.

\sb
\sd

\noindent {\bf 2. ANALYTICAL EXPRESSIONS FOR $q$-DEFORMED CGC'S}

\sb

\noindent {\bf 2.1. The Philosophy}

\sb

In this section, our aim is to derive analytical
expressions for the $U_q({\rm su}_2)$ CGc's from
a given (fully checked) formula. This can be done
with various means including:
(i) the resummation procedure,
(ii) the use of ordinary symmetry properties (corresponding to
the 12 simple ordinary symmetries of the Regge array) for the
 $U_q({\rm su}_2)$ CGc's,
(iii) the use of pure Regge symmetries
(corresponding to the
$72-12 = 60$ nonordinary symmetries of the Regge array)  for the
 $U_q({\rm su}_2)$ CGc's,
(iv) the use of the mirror reflection symmetry
     for the $U_q({\rm su}_2)$ CGc's, and
(v)  the transition from the given formula to its expression in terms of
     the $q$-deformed hypergeometric function $_3F_2$ and the use of
     symmetry properties of this function.

The resummation procedure (i) to be used below is an
adaptation, in the framework of $q$-deformations,
of the procedure described by Jucys and Bandzaitis$^8$
and applied to the standard SU$_2$ CGc's. It amounts to introduce
$$
\delta(a,b) =
\sum_s (-1)^{s-a} \;
% {q^{\pm s(a-b-1)\pm a} \over [a-s]![s-b]!}
  {q^{    s(a-b-1)+   a} \over [a-s]![s-b]!}
\eqno (5)
$$
%     (5) = ex (1.7)
in the sum occurring in a given formula for the
$U_q({\rm su}_2)$ CGc's.
The resummation of the so-obtained expression may
then be achieved by using some of the following
summation identities (or $q$-factorial sums)
$$
\sum_s \; q^{-as}\;
{1\over [s]![b-s]![c-s]![a-b-c+s]!}
= q^{-bc} \;
   {[a]!\over [b]![c]![a-b]![a-c]!}
\eqno (6)
$$
%     (6) = ex (1.3)
$$
\eqalign{
\sum_s \; (-1)^s q^{s(b+c-a-1)}\;
{[a-s]!\over [s]![b-s]![c-s]!}
= (-1)^c\; q^{bc} \;
& {[a-c]![b+c-a-1]!\over [b]![c]![b-a-1]!} , \cr
& \quad \qquad
  \quad \qquad b > a \geq c
}
\eqno (7)
$$
%     (7) = ex (1.4)
$$
\sum_s \; (-1)^s q^{s(b+c-a-1)} \;
{[a-s]!\over [s]![b-s]![c-s]!}
= q^{bc}\; {[a-b]![a-c]!\over [b]![c]![a-b-c]!} , \ a \geq b , \ a \geq c
\eqno (8)
$$
%     (8) = ex (1.5)
$$
\sum_s \; q^{s(b+c-a+2)} \;
{[b-s]![c+s]!\over s![a-s]!}
= q^{a(c+1)} \;
{[b-a]![c]![b+c+1]!\over [a]![b+c-a+1]!}
\eqno (9)
$$
%     (9) = ex (1.6)
where, as in eq.~(5), the factorials are $q$-factorials. The
identities (6)-(9) coincide with the well-known factorials
sums given in ref.~8 in the $q=1$ limit.

Among the ordinary symmetry properties (ii), we shall use the following
relations
$$(j_1j_2m_1m_2|jm)_q =
(-1)^{j_2-j-m_1} \;
q^{m_1} \;
\sqrt {{[2j+1]\over [2j_2+1]}} \;
(jj_1, - mm_1 |j_2,-m_2)_q
\eqno (10)
$$
%     (10) = ex (b)
$$(j_1j_2m_1m_2 |jm)_q =
(-1)^{j_1-j+m_2} \;
q^{-m_2}\;
\sqrt {{[2j+1]\over [2j_1+1]}}\;
(jj_2m,-m_2|j_1m_1)_q
\eqno (11)
$$
%     (11) = ex (d)
$$(j_1j_2m_1m_2|jm)_q =
(-1)^{j_1-j+m_2}\;
q^{-m_2} \;
\sqrt {{[2j+1]\over [2j_1+1]}}\;
(j_2jm_2,-m|j_1,-m_1)_q
\eqno (12)
$$
%     (12) = ex (f)
We shall also use
$$
(j_1j_2m_1m_2|jm)_q =
(-1)^{j_2+m_2}\;
q^{-m_2} \;
\sqrt {{[2j + 1]\over [2j_1+1]}}\;
(j_2j,-m_2m|j_1m_1)_{q^{-1}}
\eqno (13)
$$
%     (13) = ex (a)
$$
(j_1j_2m_1m_2|jm)_q =
(-1)^{j_1-m_1} \;
q^{m_1} \;
\sqrt {{[2j+1]\over [2j_2 + 1]}}\;
(j_1jm_1, - m|j_2,-m_2)_{q^{-1}}
\eqno (14)
$$
%     (14) = ex (c)
$$(j_1j_2m_1m_2|jm)_q =
(-1)^{j_2+m_2} \;
q^{-m_2}\;
\sqrt {{[2j+1]\over [2j_1+1]}}\;
(jj_2,-mm_2|j_1,-m_1)_{q^{-1}}
\eqno (15)
$$
%     (15) = ex (e)
where it should be observed that, in contradistinction with
   eqs.~(10)-(12), the index of the CGc in the right-hand sides
of eqs.~(13)-(15) is $q^{-1} = 1/q$.
In the limiting situation where $q=1$,
eqs.~(10)-(15) reduce to 6 of the 12
ordinary symmetry properties for the
CGc's of SU$_2$ in an
${\rm SU}_2 \supset
 {\rm U }_1$ basis.

In the following, we shall be mainly concerned with the points (i)
and (ii). However, the points (iii) (i.e., the Regge symmetries) and
(iv) (i.e., the mirror reflection symmetry: $m_1, m_2, m$ unchanged;
$j_i  \mapsto - j_i - 1$ for $i=1,2$ and
$j    \mapsto - j   - 1$)
were used to check certain forms given below.

% \sb
  \vfill\eject

\noindent {\bf 2.2. The Expressions}

\sb

The problem of finding analytical expressions  for the
$ U_q ({\rm su}_2) $   CGc's and for the corresponding
Wigner (unit tensor) operator was attacked by numerous
authors (see refs.~9  to  14  for a nonexhaustive list
of works).  Generally speaking,  the methods valid for
the ordinary CGc's  (for which $q=1$)  can be extended
to $q$-deformed CGc's.

In the limiting case where $q=1$, a useful form for the
CGc's of the nondeformed chain ${\rm su}_2 \supset {\rm u}_1$
was derived by Shapiro$^{15}$
by making use of the (L\"owdin) method of projection operators
(see also the work by Calais$^{16}$). Such a method was
adapted and applied
by Smirnov, Tolsto\u \i ~and Kharitonov$^{13}$
to the $q$-deformed chain $U_q({\rm su}_2) > {\rm u}_1$.
Our starting point is

{\it The Shapiro (Smirnov-Tolsto\u \i -Kharitonov) form (1st form)} :
$$
\eqalign{
& (j_1j_2m_1m_2 |jm)_q
=  (-1)^{j_1 + j_2 - j}
       q^{ - {1\over 2} (j_1+j_2-j)(j_1+j_2+j+1) + j_1m_2 - j_2m_1 }
\cr
&\times
 \bigg( [2j+1]
 {[j_1-j_2+j]! [j_1+j_2-j]! [j_1+j_2+j+1]! [j_2-m_2]! [j+m]!\over
  [j-j_1+j_2]! [j_2+m_2]! [j-m]! [j_1-m_1]! [j_1+m_1]!} \bigg) ^{1\over 2}
\cr
&\times
 \sum_z \,
 (-1)^{z} q^{z(j_1+m_1)}
 {[2j_2-z]![j_1+j_2-m-z]!\over
  [z]! [j_1+j_2-j-z]! [j_2-m_2-z]! [j_1+j_2+j+1-z]!}
}
\eqno (16)
$$
%     (16) = ex (1.1)
which is eq.~(5.17) of ref.~13. In eq.~(16), as well as in eqs.~(17)-(27),
it is assumed that $m = m_1 + m_2$. [If  $m \ne m_1 + m_2$,
the right-hand sides of (16)-(27) should be replaced by 0.]
The substitution $z \mapsto j_1 + j_2 - j - z$
allows us to rewrite eq.~(16) in an alternative form. We thus get

{\it The Shapiro (Smirnov-Tolsto\u \i -Kharitonov) form (2nd form)} :
$$
\eqalign{
& (j_1j_2m_1m_2 |jm)_q
= q^{ - {1\over 2}(j_1+j_2-j)( j - j_1 + j_2 - 2m_1 + 1 )+j_1m_2-j_2m_1 }
\cr
&\times
 \bigg( [2j+1]
 {[j+m]! [j_2-m_2]! [j_1-j_2+j]! [j_1+j_2-j]! [j_1+j_2+j+1]!\over
  [j-m]! [j_2+m_2]! [j_1-m_1]! [j_1+m_1]! [j-j_1+j_2]!}
 \bigg)^{1\over 2}
\cr
&\times
 \sum_z \, (-1)^z
 q^{-z(j_1+m_1)}
 {[j-m + z]! [j-j_1+j_2+z]!\over
  [z]! [2j+1+z]! [j_1+j_2-j-z]! [j-j_1-m_2+z]!}
}
\eqno (17)
$$
%     (17) = ex (1.2)
{}From the latter form, we can derive a useful intermediate form by using the
resummation procedure. Indeed, by introducing eq.~(5) into the
right-hand side of eq.~(17)
and then by using successively eq.~(8) and eq.~(7),
a long but straightforward calculation
leads to

{\it The intermediate form} :
$$
\eqalign{
& (j_1j_2m_1m_2 |jm)_q
=q^{- {1 \over 2} (j_1+j_2-j)(j_2-j_1-j+2m_2-1)+j_1m_2-j_2m_1}
\cr
&\times \bigg ([2j+1]\;
{[j_2-m_2]![j_2+m_2]![j+j_1-j_2]![j-j_1+j_2]![j_1+j_2-j]!\over
 [j_1-m_1]![j_1+m_1]![j-m]![j+m]![j+j_1+j_2+1]!}\bigg )^{1\over 2}
\cr
&\times \sum_z\, (-1)^z \;
 q^{-z(j+j_1-j_2+1)}\;
 {[j-m+z]![j_1+j_2+m-z]!\over
  [z]![j_2+m_2-z]! [j-j_1-m_2+z]! [j_1+j_2-j-z]!}
}
\eqno (18)
$$
%     (18) = ex (1.13)
which constitutes in turn an initial point for spanning other
analytical expressions of the $U_q({\rm su}_2)$ CCc's.

By starting from the intermediate form (18),
it is possible to derive,
still in the context of the resummation procedure,
the $q$-analog of the van der Waerden$^5$ (symmetrical) formula. As a
net result, we have found

{\it The van der Waerden form} :
$$
\eqalign{
& (j_1j_2m_1m_2 |jm)_q
=q^{{1\over 2} (j_1+j_2-j)(j_1+j_2+j+1)+j_1m_2-j_2m_1}
\cr
&\times \bigg ( [2j+1] \,
 {[j  - j_1 + j_2]!
  [j  + j_1 - j_2]!
  [j_1+ j_2 - j  ]!\over
  [j_1+j_2+j+1]!}\bigg )^{1\over 2}
\cr
&\times
\big ([j_1-m_1]![j_1+m_1]![j_2-m_2]![j_2+m_2]![j-m]![j+m]!\big )^{1\over 2}
\cr
&\times \sum_z  (-1)^z
q^{-z(j_1+j_2+j+1)}
{1 \over
[z]![j_1-m_1-z]![j_2+m_2-z]!}
\cr
&\times
{1 \over
[j-j_1-m_2+z]![j-j_2+m_1+z]![j_1+j_2-j-z]!}
}
\eqno (19)
$$
%     (19) = ex (2.5)
{}From the van der Waerden form (19), we can obtain the $q$-analog of
the Racah$^{6}$ formula by using the resummation procedure together
with a repeated application of the summation identity (6). This yields

{\it The Racah form (1st form)} :
$$
\eqalign{
& (j_1j_2m_1m_2 |jm)_q
= (-1)^{j_1-m_1} \>
      q^{ - {1\over 2} [j_1(j_1+1)  -
                        j_2(j_2+1)  +
                        j  (j  +1)] +
                        m_1(m  +1)  }
\cr
& \times \bigg ( [2j+1] \;
{[j_1+j_2-j]![j-m]![j+m]![j_1-m_1]![j_2-m_2]!\over
 [j+j_1-j_2]![j-j_1+j_2]![j_1+j_2+j+1]![j_1+m_1]![j_2+m_2]!}
\bigg )^{1\over 2}
\cr
& \times \sum_z \, (-1)^{z} \;
q^{ z ( j + m + 1 ) }
{[j_1+m_1+z]![j+j_2-m_1-z]!\over
 [z]![j-m-z]![j_1-m_1-z]![j_2-j+m_1+z]!}
}
\eqno (20)
$$
Then, from the Racah form (20), we can derive another useful
form, viz., the $q$-analog of the formula (13.1$||$) by
Jucys and Bandzaitis,$^8$ again through the resummation
procedure. As a matter of fact, we have obtained

{\it The Jucys-Bandzaitis form (2nd form)} :
$$
\eqalign{
& (j_1j_2m_1m_2 |jm)_q
= (-1)^{j_1-m_1}
q^{-{1\over 2} [j_1(j_1 + 1)
               -j_2(j_2 + 1)
               +j  (j   + 1)] - j_1j + j_1m + jm_1 + m_1}
\cr
&\times \bigg (
{[j+j_1+j_2+1]! \over
 [-j+j_1+j_2]![j-j_1+j_2]![j+j_1-j_2]!}\bigg )^{1\over 2}
\cr
&\times \bigg ( [2j+1]
{[j_1-m_1]![j_1+m_1]![j-m]![j+m]!\over
[j_2-m_2]![j_2+m_2]!}\bigg )^{1\over 2}
\cr
&\times \sum_z(-1)^zq^{z(j_1-j_2+j)}
{[j_2+j-m_1-z]![j_1+j_2-m-z]!\over
 z![j_1-m_1-z]![j-m-z]![j_1+j_2+j+1-z]!}
}
\eqno (21)
$$
Going back to the van der Waerden form,
by making the substitution $z \mapsto j_2 + m_2 - z$ in eq.~(19)
and by applying the resummation procedure to the so-obtained relation,
we arrive at

{\it The Majumdar form (1st form)} :
$$
\eqalign{
& (j_1j_2m_1m_2 |jm)_q
=(-1)^{j_2+m_2} \;
  q^{ - {1\over 2} (j - j_1 + j_2) (j_1 + j_2 - j + 1) + m_1j - mj_1 - m_2 }
\cr
&\times \bigg ( [2j+1] \;
 {[j-j_1+j_2]![j_1-m_1]![j_1+m_1]![j_2-m_2]![j+m]!\over
  [j+j_1-j_2]![j_1+j_2-j]![j_1+j_2+j+1]![j_2+m_2]![j-m]!}\bigg )^{1\over 2}
\cr
&\times \sum_z \, (-1)^z \;
 q^{z (j_1-m_1+1) }\;
 {[2j-z]![j_1+j_2-j+z]!\over
  [z]![j+m-z]![j_1-j-m_2+z]![j-j_1+j_2-z]!}
}
\eqno (22)
$$
%     (22) = ex (M1) = ex (3.5)
Finally, the substitution $z \mapsto j + m - z$ in eq.~(22) produces

{\it The Majumdar form (2nd form)} :
$$
\eqalign{
& (j_1j_2m_1m_2 |jm)_q
    =(-1)^{j-j_2+m_1}
     q^{ {1\over 2} [j_1(j_1 + 1)
                    -j_2(j_2 + 1)
                    +j  (j   + 1)] - m_1(m - 1) }
\cr
&\times \bigg ( [2j+1] \;
 {[j-j_1+j_2]![j_1-m_1]![j_1+m_1]![j_2-m_2]![j+m]!\over
  [j+j_1-j_2]![j_1+j_2-j]![j_1+j_2+j+1]![j_2+m_2]![j-m]!}\bigg )^{1\over 2}
\cr
&\times \sum_z \, (-1)^{z} \;
q^{ - z (j_1 - m_1 + 1) } \;
{[j-m+z]![j_1+j_2+m-z]!\over
 [z]![j+m-z]![j_1+m_1-z]![j_2-j_1-m+z]!}
}
\eqno (23)
$$
Equation  (22)  is the $q$-analog of a formula
originally  derived  by  Majumdar$^{7}$
while eq.~(23) is a simple consequence of (22).

Other analytical formulas for the CGc's of $U_q({\rm su}_2) > {\rm u}_1$
may be spanned from the just obtained forms owing to the symmetry
properties (10-15). For example, by applying the symmetry
property (13) to the intermediate form (18) we obtain

{\it The Wigner form} :
$$
\eqalign{
& (j_1j_2m_1m_2 |jm)_q
= (-1)^{j_2+m_2} \;
   q^{{1\over 2} (j-j_1+j_2) (j-j_1-j_2+2m-1)-j_2m-(j+1)m_2}
\cr
&\times \bigg( [2j+1]
 {[j_1+j_2-j]! [j-j_1+j_2]![j+j_1-j_2]![j-m]![j+m]!\over
  [j_1+j_2+j+1]![j_1-m_1]![j_1+m_1]![j_2-m_2]![j_2+m_2]!}\bigg )^{1\over 2}
\cr
&\times \sum_z \, (-1)^z\;
 q^{z(j_1+j_2-j+1)}
 {[j_1-m_1+z]![j+j_2+m_1-z]!\over
  [z]![j+m-z]![j_1-j_2-m+z]![j-j_1+j_2-z]!}
}
\eqno (24)
$$
[In other words, by introducing the intermediate form (18) for the CGc
in the right-hand side of (13),  then the Wigner form (24) is obtained
as the  left-hand side of (13).]
Similarly,
the introduction of the intermediate form (18) in the
right-hand side of the symmetry property (11) generates

{\it The Zukauskas-Mauza form} :
$$
\eqalign{
& (j_1j_2m_1m_2 |jm)_q
=(-1)^{j_1-j+m_2}\;
 q^{-{1\over 2} (j-j_1+j_2)(j_2-j_1-j-2m_2-1)-(j+1)m_2-j_2m}
\cr
&\times \bigg( [2j+1]
 {[j_1+j_2-j]![j-j_1+j_2]![j+j_1-j_2]![j_2-m_2]![j_2+m_2]!\over
  [j_1+j_2+j+1]![j_1-m_1]![j_1+m_1]![j-m]![j+m]!}\bigg)^{1\over 2}
\cr
&\times \sum_z \, (-1)^z \;
 q^{-z(j_1-j_2+j+1)}
{[j_1-m_1+z]![j+j_2+m_1-z]!\over
 [z]![j_2-m_2-z]![j_1-j+m_2+z]![j-j_1+j_2-z]!}
}
\eqno (25)
$$
The $q$-analog of the Racah form used by Kibler and Grenet in ref.~1
[at the level of their eq.~(92)]
can be deduced from the Racah form (20) : it is sufficient to introduce
(20) into the symmetry property (14); this leads to

{\it The Racah form (2nd form)} :
$$
\eqalign{
& (j_1j_2m_1m_2 |jm)_q
=q^{{1\over 2} [j_1(j_1+1)
                +j_2(j_2+1)
                -j  (j  +1)] + m_1(m_2 - 1)}
\cr
&\times \bigg ( [2j+1] \;
 {[j+j_1-j_2]![j_1-m_1]![j_2-m_2]![j_2+m_2]![j+m]!\over
  [j_1+j_2-j]![j-j_1+j_2]![j_1+j_2+j+1]![j_1+m_1]![j-m]!}\bigg)^{1\over 2}
\cr
&\times \sum_z \, (-1)^z q^{ - z (j_2 - m_2 + 1) }\;
 {[j_1+m_1+z]![j+j_2-m_1-z]!\over
  [z]![j_1-m_1-z]![j_2+m_2-z]![j-j_2+m_1+z]!}
}
\eqno (26)
$$
Finally, by putting the Shapiro form
(1st form) [eq.~(16)] in the right-hand side
of the symmetry property (12), we obtain

{\it The Jucys-Bandzaitis form (1st form)} :
$$
\eqalign{
& (j_1j_2m_1m_2 |jm)_q =
\; (-1)^{j_2+m_2}
 \; q^{{1\over 2}(j_1-j_2-j)(j_1+j_2+j+1) - mj_2 - m_2j - m_2}
\cr
&\times \bigg (
{[-j_1+j_2+j]![j_1+j_2-j]![j_1+j_2+j+1]!\over
 [ j_1-j_2+j]!} \bigg )^{1\over 2}
\cr
&\times \bigg (
[2j+1] \;
{[j_1-m_1]![j+m]!\over
 [j_1+m_1]![j_2-m_2]![j_2+m_2]![j-m]!}\bigg )^{1\over 2}
\cr
&\times \sum_z \; (-1)^z \; q^{z(j_2+m_2)}\;
{[2j-z]![j+j_2+m_1-z]!\over [z]![j+m-z]![j+j_2-j_1-z]![j_1+j_2+j+1-z]!}
}
\eqno (27)
$$
which is the $q$-analog of the relation (13.1B) obtained by
Jucys and Bandzaitis and listed in ref.~8.

\sb
\sd

\noindent {\bf 3. TOWARDS $q$-BOSON REALIZATIONS OF THE $U_q({\rm su}_2)$
CGC'S}

\sb

\noindent {\bf 3.1. The Philosophy}

\sb

Following the approach of ref.~1, we are now in a position to
find $q$-boson realizations of the $U_{q}({\rm su}_2)$ unit
tensor. The components $t[q : k \rho \Delta]$
of such a tensor operator are defined by$^{1,2}$
$$
\langle j' m' \vert t[q : k \rho \Delta] \vert j m
\rangle :=
\delta(j', j + \Delta)
\delta(m', m + \rho  ) (-1)^{2k}
\left ( [2j'+1]_q \right )^{-{1\over2}}
                (j k m \rho \vert j' m')_q
\eqno (28)
$$
where $(j k m \rho \vert j' m')_q$ is a CGc for $U_{q}({\rm su}_2)$.
The operator $t[q : k \rho \Delta]$
constitutes a $q$-deformation of the operator
$t_{k q \alpha} \equiv t[1 : k q \alpha]$
worked out by Kibler and Grenet.$^1$

It was shown in ref.~1 that the operator
$t[1 : k q \alpha]$ can be expressed in the
enveloping algebra of the Schwinger
algebra (isomorphic with so$_{2,3}$). The
extension to the case where $q \ne 1$ is trivial so
that it is possible
to find a realization of the operator $t[q : k \rho \Delta]$
in terms of $q$-boson operators defined in a
two-particle Fock space (with two sets
$\left\{ a_+, a_+^+ \right\}$ and
$\left\{ a_-, a_-^+ \right\}$ of $q$-bosons). As in ref.~2,
we take Macfarlane$^{17}$ and Biedenharn$^{18}$ $q$-bosons corresponding to
$$
\eqalign{
  a_+   \; |n_1n_2 \rangle \; =
        \; &{\sqrt {[n_1    ]_q}} \; |n_1 - 1, n_2 \rangle \cr
  a_+^+ \; |n_1n_2 \rangle \; =
        \; &{\sqrt {[n_1 + 1]_q}} \; |n_1 + 1, n_2 \rangle \cr
  a_-   \; |n_1n_2 \rangle \; =
        \; &{\sqrt {[n_2    ]_q}} \; |n_1, n_2 - 1 \rangle \cr
  a_-^+ \; |n_1n_2 \rangle \; =
        \; &{\sqrt {[n_2 + 1]_q}} \; |n_1, n_2 + 1 \rangle \cr
       N_i |n_1n_2 \rangle \; = \; &n_i |n_1n_2 \rangle \quad (i=1,2)
}
\eqno (29)
$$
[from which it is clear that
$a_+^+ = (a_+)^{\dagger}$ and
$a_-^+ = (a_-)^{\dagger}$ when $q \in \gr$ or   $q \in S^1$].
Then, from the transformation
$$
     |jm \rangle \, \equiv \, |n_1n_2 \rangle      \qquad
             j \, := \, {1 \over 2} (n_1 + n_2)     \qquad
             m \, := \, {1 \over 2} (n_1 - n_2)
\eqno (30)
$$
we have
$$
\eqalign{
a_{\pm}   \; |jm \rangle \; = \; & {\sqrt{[j \pm m    ]_q}} \;
|j - {1\over 2}, m \mp {1\over 2} \rangle \cr
a_{\pm}^+ \; |jm \rangle \; = \; & {\sqrt{[j \pm m + 1]_q}} \;
|j + {1\over 2}, m \pm {1\over 2} \rangle
}
\eqno (31)
$$
which is at the root of the (Jordan-Schwinger) realization
of $U_{q}({\rm su}_2)$. A simple iteration of (31) yields
$$
\eqalign{
\left ( a_{+}\right )^n |jm \rangle \; & = \;
 \left (
 {\displaystyle [j + m    ]_q!\over
  \displaystyle [j + m - n]_q!} \right )^{1/2}
\big | j - {n\over 2} , m - {n\over 2} \rangle
\cr
\left ( a_{-}\right )^n |jm \rangle \; & = \;
 \left (
 {\displaystyle [j - m    ]_q!\over
  \displaystyle [j - m - n]_q!} \right )^{1/2}
\big | j - {n\over 2} , m + {n\over 2} \rangle
\cr
\left ( a^+_+ \right )^n |jm \rangle \; & = \;
 \left (
 {\displaystyle [j + m + n]_q!\over
  \displaystyle [j + m    ]_q!} \right )^{1/2}
\big | j + {n \over 2}, m + {n\over 2} \rangle
\cr
\left ( a^+_- \right )^n |jm \rangle \; & = \;
 \left (
 {\displaystyle [j - m + n]_q!\over
  \displaystyle [j - m    ]_q!} \right )^{1/2}
\big | j + {n \over 2}, m - {n\over 2} \rangle
}
\eqno (32)
$$
for $n \in \grn$. Thus, it is possible to understand
why $t[q : k  \rho  \Delta]$ can be developed in the
polynomial form
$$
t[q : k  \rho  \Delta]
= \sum_{\alpha, \beta, \gamma, \delta}
     C_{\alpha  \beta  \gamma  \delta} (q,k,\rho,\Delta)
 \left ( a^+_+ \right )^\alpha
 \left ( a_+   \right )^\beta
 \left ( a^+_- \right )^\gamma
 \left ( a_-   \right )^\delta
\eqno (33)
$$
where the sum on $\alpha$, $\beta$, $\gamma$, and $\delta$
is in fact a sum on a single variable $z$ [likewise in eqs.~(16)-(27)].

\sb

\noindent {\bf 3.2. Some $q$-Bosonized Expressions}

\sb

As an example, the $q$-boson realization of the operator
$t[ q : k \rho \Delta ]$ corresponding to eq.~(18) gives

{\it The intermediate realization} :
$$
\eqalign{
t[ q : k \rho \Delta ] \;
& = \; (-1)^{ 2k } \;
  q^{ {1\over 2} (k - \Delta)(2j + \Delta - k + 1 - 2  \rho) + j \rho - km }
\cr
& \times \; \left (
{ \displaystyle
  [k - \Delta ]!
  [k + \Delta ]!
  [k - \rho   ]!
  [k + \rho   ]! [2j + \Delta - k]!\over
  \displaystyle
  [2j + \Delta + k + 1 ]! } \right )^{1 \over 2}
\cr
& \times \; \sum_z \; (-1)^{ z } \;
  q^{ -z(2j + \Delta - k + 1) }\;
  {\displaystyle (a_+)^{k-\Delta - z} (a^+_+)^{k+\rho-z}
                 (a_-)^z              (a^+_-)^{\Delta-\rho +z}\over
   \displaystyle [k-\Delta-z     ]! \;
                 [k+\rho-z       ]! \;
                 [z              ]! \;
                 [\Delta - \rho+z]!}
}
\eqno (34)
$$
where, as in section 2, the abbreviation $[x]$ stands for $[x]_q$.
The correctness of eq.~(34) can be verified by taking the
$j'm'$-$jm$ matrix element of the right-hand side of (34):
then, by using (32) and (28), it can be checked that we obtain the CGc
                $(j k m \rho \vert j' m')_q$
in the intermediate form (18).
It is to be observed that, in equations of type (34),
the $q$-factorials in the
denominators of the sum  over  $z$  give a guaranty
that the powers  of  all  $q$-boson  operators  are
nonnegative integers.

In a similar way, we have obtained the $q$-analogs
of eqs.~(89)-(92) of ref.~1. They are given by

{\it The van der Waerden realization} :
$$
\eqalign{
t&[q : k \rho \Delta] \; = \;
 (-1)^{k + \Delta} \;
q^{ {1 \over 2} (k - \Delta) (2j + \Delta + k + 1) + j \rho - km}
\cr
 & \times
   \left ( { \displaystyle [k + \rho ]!
                           [k - \rho ]!
                           [k + \Delta]!
                           [k - \Delta]! [2j + \Delta - k]!\over
         \displaystyle  [2j + \Delta + k + 1] !} \right )^{1/2}
\cr
 & \times \;
  \sum_z \; (-1)^z \; q^{ -z (2j + \Delta + k + 1) } \;
 {\displaystyle (a^+_-)^{k - \rho - z} (a_-)^{k - \Delta - z}
                (a^+_+)^{\rho + \Delta + z} (a_+)^z\over
  \displaystyle  [k- \rho -z]! \; [k- \Delta - z]! \;
                 [ \rho + \Delta + z]! \; [z] !}
}
\eqno (35)
$$

{\it The Zukauskas-Mauza realization} :
$$
\eqalign{
t&[q : k \rho \Delta] \; = \;
 (-1)^{ \rho + \Delta} \;
q^{ {1 \over 2} (k + \Delta) (2j + \Delta - k +1 + 2 \rho)
                            -( j + \Delta + k +1)\rho - km }
\cr
 & \times
  \left ( { \displaystyle [k + \rho ]!
                          [k - \rho ]!
                          [k + \Delta ]!
                          [k - \Delta ]! [2j + \Delta - k]!\over
            \displaystyle [2j + \Delta + k + 1]!} \right )^{1/2}
\cr
& \times \;
  \sum_z \; (-1)^z \; q^{ - z (2j + \Delta - k + 1) } \;
 {\displaystyle (a_+)^{k - \rho - z} (a^+_+)^{k + \Delta - z}
                (a_-)^{ \rho - \Delta + z} (a^+_-)^z \over
  \displaystyle [k - \rho - z]! \; [k + \Delta - z]! \;
                [ \rho - \Delta + z]! \; [z] !}
}
\eqno (36)
$$

{\it The Wigner realization} :
$$
\eqalign{
t&[q : k \rho \Delta] \; = \;
 (-1)^{k- \rho } \;
q^{ - {1 \over 2} (k + \Delta) (\Delta + k +1 - 2 m - 2 \rho)
                            -( j + \Delta + k +1)\rho - km  }
\cr
& \times
  \left ( {\displaystyle [k + \Delta ]!
                         [k - \Delta ]! [2j + \Delta - k]!\over
           \displaystyle [k + \rho ]!
                         [k - \rho ]!   [2j + \Delta + k + 1]!} \right )^{1/2}
\cr
& \times \;
  \sum_z \; (-1)^z \; q^{ z (k - \Delta + 1) } \;
 {\displaystyle (a^+_+)^z                (a_+)^{k - \rho }
                (a^+_+)^{k + \Delta - z}
                (a^+_-)^{k + \Delta - z} (a_-)^{k + \rho } (a^+_-)^z\over
  \displaystyle [k + \Delta - z]! \; [z]!}
}
\eqno (37)
$$

{\it The Racah (2nd) realization} :
$$
\eqalign{
t&[q : k \rho \Delta] \; = \;
 (- 1)^{2k} \;
q^{ {1 \over 2} [k(k + 1) - \Delta (2j + \Delta + 1)] + m (\rho - 1) }
\cr
 & \times
   \left ( {\displaystyle [k + \rho ]!
                          [k - \rho ]! [2j + \Delta - k]!\over
            \displaystyle [k + \Delta ]!
                          [k - \Delta ]!
  [2j +  \Delta + k + 1]!} \right )^{1/2}
\cr
 & \times \;
  \sum_z \; (- 1)^z \; q^{ - z (k - \rho + 1) } \;
 {\displaystyle (a^+_+)^{k + \rho -z} (a_+)^{k - \Delta}
                (a^+_+)^z             (a_-)^{k + \rho - z}
                (a_-^+)^{k + \Delta}  (a_-)^z\over
  \displaystyle [k + \rho -z]! \; [z]!}
}
\eqno (38)
$$
Note that in eqs.~(34)-(38), it may be appropriate to
replace the eigenvalues $j$ and $m$ {\it in terms of}
               the operators $(1/2)(N_1 + N_2)$ and
                             $(1/2)(N_1 - N_2)$, respectively,
in order to have basis independent operators.

The remaining forms of section 2 can be $q$-bosonized
too. The detailed results shall be given elsewhere. Remark
that the Hermitean conjugation property
$$
t[q : k  \rho  \Delta] ^{\dagger} = (-1)^{\alpha - \rho}
t[q : k -\rho -\Delta]
\eqno (39)
$$
[that is connected with the permutation of $j$ and $j'$ in eq.~(28)]
may be exploited to pass from one realization to another
or to produce other $q$-boson realizations.

\sb
\sd

\noindent {\bf 4. EXTENSION TO THE QUANTUM ALGEBRA $U_{qp}({\rm u}_2)$}

\sb

The question of extending the one-parameter algebra $U_{q}({\rm su}_2)$
                         to a two-parameter algebra was addressed by
several authors.$^{19-30}$ Indeed,
the $qp$-quantized universal enveloping algebra $U_{qp}({\rm su}_2)$
can be seen to be amenable to the one-parameter algebra
 $U_{q}({\rm su}_2)$. To get a truly two-parameter algebra, it is necessary to
$qp$-deform  u$_2$ rather than
            su$_2$. We follow here the presentation of ref.~27
(see also ref.~2): the two-parameter
quantum algebra $U_{qp}({\rm u}_2)$ is spanned by the four generators
$J_{\alpha}$ (with $\alpha = 0,3,+,-$) which satisfy the following
commutation relations
 $$
[J_3,J_\pm   ] \ = \ \pm J_\pm                      \qquad
[J_+,J_-     ] \ = \ (qp)^{J_0-J_3} \ [[2J_3]]_{qp} \qquad
[J_0,J_\alpha] \ = \ 0, \ \alpha=0,3,+,-
\eqno (40)
 $$
[Observe that the two first commutation
relations in (40) reduce to (4)
when $p=q^{-1}$.] In order to endow $U_{ {qp}}({\rm u}_2)$ with a Hopf
algebraic
structure, it is necessary to introduce a co-product $\Delta_{qp}$.
It is defined by the application
$\Delta_{qp} \ : \ U_{ {qp}}({\rm u}_2) \otimes
                   U_{ {qp}}({\rm u}_2) \to
                   U_{ {qp}}({\rm u}_2)$
such that
$$
\eqalign{
\Delta_{qp}(J_0)     & := J_0 \> \otimes \> 1  +  1 \> \otimes \> J_0 \cr
\Delta_{qp}(J_3)     & := J_3 \> \otimes \> 1  +  1 \> \otimes \> J_3 \cr
\Delta_{qp}(J_{\pm}) & := J_{\pm} \> \otimes
\> {(qp)}^{{{1}\over{2}}J_0} \> {(qp^{-1})}^{+{{1}\over{2}}J_3} +
   {(qp)}^{{{1}\over{2}}J_0} \> {(qp^{-1})}^{-{{1}\over{2}}J_3} \>
\otimes \> J_{\pm}
}
\eqno (41)
$$
[Note that with the constraint $p = q^{\ast}$, the co-product
satisfies the Hermitean conjugation property
 $(\Delta_{qp} (J_\pm))^\dagger = \Delta_{pq}(J_\mp)$
and is compatible with the commutation relations for the four operators
$\Delta_{qp} (J_\alpha)$ with $\alpha = 0,3,+,-$.]
The universal ${\cal{R}}$-matrix associated
to the co-product $\Delta_{qp}$ reads
$$
{\cal{R}}_{pq} = \pmatrix{
p&0&0&0          \cr
0&\sqrt{pq}&0&0  \cr
0&p-q&\sqrt{pq}&0\cr
0&0&0&p
}
\eqno (42)
$$
and it can be proved that ${\cal{R}}_{pq}$ verifies the so-called
Yang-Baxter equation. The co-unit and anti-pode required for the
Hopf algebraic structure of $U_{ {qp}}({\rm u}_2)$ are given in
ref.~27.

The operator defined by$^{27}$
  $$
 C_2( U_{ {qp}}({\rm u}_2))\  :=  \  {{1}\over{2}}(J_+J_- + J_-J_+)\  +\
 {{1}\over{2}} \> [[2  ]]_{qp} \> (qp)^{J_0-J_3} \>
           \left( [[J_3]]_{qp} \right)^2
\eqno (43)
$$
 is an invariant of the quantum algebra $U_{ {qp}}({\rm u}_2)$. The
latter
invariant gives back the well-known invariant of the quantum algebra
 $U_{ {q}}({\rm su}_2)$ when $p=q^{-1}$.

 In the case
where neither $q$ nor $p$ are roots of unity, the
representation theory of $U_{{qp}}({\rm u}_2)$
easily follows from the one of the Lie algebra u$_2$.
 An irreducible
representation of the quantum algebra
 $U_{{qp}}({\rm u}_2)$
is characterized by a doublet ($j_0,j$) where $j_0 \in \gr$ and $2j \in \grn$.
Such a representation is associated to a subspace
${\cal E}(j_0,j) =
\left\{ \vert j_0 j m \rangle \ : \ m = -j, -j+1, \cdots, j \right\}$.
The generic basis vector
$\vert j_0 j m \rangle$ of
${\cal E}(j_0,j)$ can be obtained from the highest weight vector
$\vert j_0 j j \rangle$ owing to
$$
\vert j_0 j m \rangle \; = \; (qp)^{-{1 \over 2}(j_0-j)(j-m)}
\, \sqrt {{[[j+m]]_{qp}! \over [[2j]]_{qp}! [[j-m]]_{qp}!}}
\, (J_-)^{j-m}
\, \vert j_0 j j \rangle
\eqno (44)
$$
Then, the action of the generators $J_{\alpha}$ (with $\alpha = 0,3,+,-$)
on the subspace ${\cal E}(j_0,j)$ is given by
$$
\eqalign{
J_0 \, | j_0 j m \rangle &=           j_0 \, |j_0 j m   \rangle \cr
J_3 \, | j_0 j m \rangle &=           m   \, |j_0 j m   \rangle \cr
J_+ \, | j_0 j m \rangle &= (qp)^{{1 \over 2}(j_0-j)}
\sqrt { [[j-m]]_{qp}  [[j+m+1]]_{qp} }    \, |j_0 j m+1 \rangle \cr
J_- \, | j_0 j m \rangle &= (qp)^{{1 \over 2}(j_0-j)}
\sqrt { [[j+m]]_{qp}  [[j-m+1]]_{qp} }    \, |j_0 j m-1 \rangle
}
\eqno (45)
$$
The eigenvalues of the invariant operator
$C_2(U_{{qp}}({\rm u}_2))$ on ${\cal E}(j_0,j)$ read
$$
{ {q^{j + j_0 + 1} p^{    j_0 - j}
 - q^{    j_0 + 1} p^{    j_0    }
 - q^{    j_0    } p^{    j_0 + 1}
 + q^{    j_0 - j} p^{j + j_0 + 1} } \over {(q-p)^2} } \, = \,
(qp)^{(j_0-j)} \, [[j]]_{qp} \, [[j+1]]_{qp}
\eqno (46)
$$
in the general case where $j_0 \ne j$. In the particular case where
 $j_0 = j$, the eigenvalues (46) are equal to $[[j]]_{qp} \, [[j+1]]_{qp}$,
a fact that was used as a basic ingredient
for the $qp$-rotor model developed in refs.~31 and 32.

The quantum algebra $U_{{qp}}({\rm u}_2)$
clearly depends on the two parameters $q$ and $p$.
It is however interesting to
show its relation with the well-known one-parameter algebra
$U_{q}({\rm su}_2)$.
In this respect, eq.~(41)
suggests the following change of parameters
$$
Q := (qp^{-1})^{1\over 2} \qquad \quad
P := (qp)     ^{1\over 2}
\eqno (47)
$$
In terms of the parameters $Q$ and $P$, we have
$$
[[x]]_{qp}  = P^{x-1}                 \, [x]_Q \qquad \quad
[[x]]_{qp}! = P^{{1 \over 2} x (x-1)} \, [x]_Q!
\eqno (48)
$$
Then, by introducing the generators $A_\alpha$ (with $\alpha = 0,3,+,-$)
$$
  A_0   \, := \, J_0 \qquad \quad
  A_3   \, := \, J_3 \qquad \quad
  A_\pm \, := \, (qp)^{-{1\over 2}(J_0 - {1\over 2})} \, J_\pm
\eqno (49)
$$
it can be shown that the two-parameter quantum algebra
 $U_{{qp}}({\rm u}_2)$ is isomorphic to the central
extension
$$
U_{qp}({\rm u}_2) = {\rm u}_1 \otimes U_Q({\rm su}_2)
\eqno (50)
$$
where $ {\rm u}_1$ is spanned by the operator $A_0$ and $U_Q({\rm su}_2)$
by the set $\{A_3, A_+, A_-\}$. The $Q$-deformation   $U_Q({\rm su}_2)$
of the Lie algebra  ${\rm su}_2$ corresponds to the usual commutation
relations [cf.~eq.~(4)]
 $$
[A_3,A_\pm   ] \ = \ \pm A_\pm \qquad \quad
[A_+,A_-     ] \ = \ [2A_3]_Q
\eqno (51)
$$
(Of course, we have $[A_0,A_{\alpha}]=0$ for $\alpha = 0, 3, +, -$.)
Furthermore, the co-product relations (41) leads to
 $$
\eqalign{
\Delta_{qp}(J_{\pm}) \> & = \> \left(  P^{ A_0 - { {1} \over {2} } } \otimes
                                       P^{ A_0                     } \right)
\> \Delta_{Q}(A_{\pm}) \cr
                        & = \> P^{ \Delta_{Q}(A_0) - { {1} \over {2} } }
\> \Delta_{Q}(A_{\pm})
}
\eqno (52)
$$
where the co-product
$\Delta_Q \ : \ U_Q({\rm su}_2) \otimes
                U_Q({\rm su}_2) \to
                U_Q({\rm su}_2)$
is given via
$$
\eqalign{
\Delta_{Q}(A_0)     & := A_0 \> \otimes \> 1 + 1 \> \otimes \> A_0 \cr
\Delta_{Q}(A_3)     & := A_3 \> \otimes \> 1 + 1 \> \otimes \> A_3 \cr
\Delta_{Q}(A_{\pm}) & := A_{\pm} \> \otimes \> Q^{+A_3} +
                                              Q^{-A_3}
                                \> \otimes \> A_{\pm}
}
\eqno (53)
$$
The invariant  $ C_2( U_{ {qp}}({\rm u}_2))$ may be
transcribed in terms of the two parameters $Q$ and $P$. In fact,
eq.~(43) can be rewritten as
$$
  C_2(U_{qp}({\rm u}_2)) \; = \;
  P^{2A_0-1} \; C_2(U_Q({\rm su}_2))
\eqno (54)
$$
where
$$
C_2(U_Q({\rm su}_2)) \; := \;
  {1 \over 2} \; (A_+A_- + A_-A_+) +
  {1 \over 2} \; [2]_Q \; \left( [A_3]_Q \right)^2
\eqno (55)
$$
so that the eigenvalues (46) of  $ C_2( U_{ {qp}}({\rm u}_2))$
on the subspace ${\cal E}(j_0,j)$ can be rewritten as
$
P^{2j_0 - 1}\, [j]_Q \, [j+1]_Q
$.

{}From the point of view of the representation theory, eqs.~(47)-(55) strongly
suggest that the coupling
(i.e., CGc's and $3-jm$ symbols) and recoupling
(i.e., $6-j$ and $9-j$ symbols) coefficients of
 $ U_{ {qp}}({\rm  u}_2) $ may be deduced from the corresponding ones of
 $ U_{ {Q }}({\rm su}_2) $. This intuition is further reinforced by the
fact that the CGc's
$$
             (j_{01} j_{02} j_1 j_2  m_1 m_2 \vert j_0 j m )_{qp}
            \, \equiv \,
                                    (m_1 m_2 \vert       m )_{qp}
\eqno (56)
$$
of  $ U_{ {qp}}({\rm  u}_2) $ are easily seen to satisfy the three-term
recursion relations
$$
\eqalign
{
 \sqrt{ [j \mp m ]_Q \ [j \pm  m + 1) ]_Q } \ &
        (m_1        m_2       \vert m \pm 1)_{qp} \cr
 = & \ Q^{+ m_2} \ \sqrt{ [j_1 \pm m_1]_Q \ [j_1 \mp m_1 + 1]_Q } \
        (m_1 \mp 1, m_2       \vert m      )_{qp} \cr
 + & \ Q^{- m_1} \ \sqrt{ [j_2 \pm m_2]_Q \ [j_2 \mp m_2 + 1]_Q } \
        (m_1,       m_2 \mp 1 \vert m      )_{qp}
}
\eqno (57)
$$
which are identical to the ones satisfied by the CGc's
   $(j_1 j_2 m_1 m_2 \vert     j  m )_{Q}$
of the quantum algebra  $ U_{ {Q}}({\rm su}_2) $ (see ref.~3).
Therefore, there exists a proportionality relation
between the $qp$-CGc's and the
             $Q$-CGc's.    Reality and normalization conditions can
be used to justify that the proportionality constant is equal to 1.
Indeed, this may be checked by direct calculation:
by adapting to the $qp$-deformation  $ U_{ {qp}}({\rm  u}_2) $ the
method of projection operators used for su$_2$ (in ref.~15)
and         for       $ U_{ {q}}({\rm su}_2) $ (in ref.~13), we can show that
we have the connecting formula:
$$
(j_{01}j_{02}j_1j_2m_1m_2|j_0jm)_{qp} \; = \; \delta(j_0, j_{01} + j_{02}) \;
            (j_1j_2m_1m_2|   jm)_{Q }
\eqno (58)
$$
a result to be compared with the ones in refs.~28 and 29 for
$ U_{ {qp}}({\rm su}_2) $. Therefore,
all formulas of subsection 2.2 may be adapted to the case of the
two-parameter quantum algebra $ U_{ {qp}}({\rm u}_2) $.

\sb
\sd

\noindent {\bf 5. CLOSING REMARKS}

\sb

We obtained in this work various analytical forms
for the $U_q({\rm su}_2)$ CGc's.
A special effort was put on the compatibility between
the different forms discussed. Indeed, all the obtained
forms were deduced from one single form, namely, the so-called
Shapiro form,
 either by
applying the resummation procedure or by using ordinary symmetry
properties of the CGc's.
Some other checks (not reported in the present paper) were also
achieved by means of Regge symmetries and of the mirror reflection
symmetry extended to the $U_q({\rm su}_2) > {\rm u}_1$  chain.
In addition, the forms were compared to existing formulas when possible.

In the classical limit where $q=1$, the various $q$-dependent
forms reduce to well-known expressions
(like the       formulas derived by Wigner, van der Waerden, and Racah
in the early days of what is refered to as Wigner-Racah algebra
of the rotation group)
          and to less-known expressions
for the CGc's of the group
SU$_2$ in an
SU$_2 \supset {\rm U}_1$ basis.
For a given form, the passage from $q = 1$ to $q \ne 1$ manifests
itself by: (i) the replacement of $ 2j+1   $ by
                                  $[2j+1]_q$ and of ordinary factorials
                                                     by $q$-factorials,
(ii)  the introduction of an {\it internal} $q$-factor which depends on a
summation index $z$, and
(iii) the introduction of an {\it external} $q$-factor ($z$-independent).

It is remarkable that the internal $q$-factor (which is $z$-dependent)
assumes the
form $q^{\pm z(N - D + 1)}$, where $N$ and $D$ stand respectively
for the numerator and the
denominator of the fraction involved in the sum over $z$. The occurrence of
this heuristic rule might be rationalized on the basis of some properties of
the $q$-deformation of the hypergeometric function $_3F_2$ in term of which it
is possible to express the $q$-deformed CGc's of SU$_2$ [see ref.~33
for the use of $_3F_2(abc,de;1)$ in connection with SU$_2$ CGc's].

The various forms of the
$ U_{ {q}}({\rm su}_2)$  CGc's
fall in three families:
the Wigner, van der Waerden, and
Racah families. Each family is characterized by a given distribution of the
$q$-factorials $[j_1 \pm m_1]_q!$,
               $[j_2 \pm m_2]_q!$, and
               $[j   \pm m  ]_q!$
in the numerator of the factor in front of the sum over $z$.
The van der Waerden family has the six $q$-factorials in the numerator.
The          Wigner family
[including the Shapiro forms,
      the intermediate form,
   the Zukauskas-Mauza form, and
the Jucys-Bandzaitis (1st) form]
presents two $q$-factorials in the numerator
while the Racah family
[including the Majumdar forms and the
the Jucys-Bandzaitis (2nd) form]
has four $q$-factorials in the numerator.
In principle, the parents in a family may be obtained from
any member of the family
by applying ordinary symmetry properties of the CGc's.

Some preliminary results on the $q$-bosonization of the CGc's of
$U_{q}({\rm su}_2)$ have been reported. Various expressions
have been given for the unit tensor operator $t[q: k \rho \Delta]$.
Not all the possible forms of the operator $t[q: k \rho \Delta]$
have been described. In particular, the Majumdar realization of
$t[q: k \rho \Delta]$ has been omitted since it presents some
peculiarities as in the $q=1$ case.  A more complete listing
of $q$-boson realizations of the operator $t[q: k \rho \Delta]$
shall be published elsewhere.

Finally, we have extended to a two-parameter
deformation of u$_2$ the various
expressions for the $q$-deformed CGc's of the chain
${\rm SU}_2 \supset {\rm U}_1$.
The derivation of a
true (nontrivial) two-parameter deformation of su$_2$ has been questioned
by various authors. Although, the CGc's for
$U_{qp}({\rm  u}_2)$ may be obtained from the ones for
 $U_{Q}({\rm su}_2)$ by means of a formula where the
two parameters $q$ and $p$ are unified in a single one [i.e.,
$Q= (qp ^{-1} ) ^ {1 \over 2}$], the two parameters $q$ and $p$
                                  [or, equivalently, $Q$ and
$P= (qp       ) ^ {1 \over 2}$] really
appear in the Casimir(s) of the Hopf algebra
$U_{qp}({\rm  u}_2)$. For the practitioner,
interested in putting some numbers on the (real) world,
the two parameters $q$ and $p$ may have some interest
when dealing with a comparison between theory and experiment (via
fitting procedures for example).  The latter point was applied to
the derivation of a $qp$-rotor model for describing rotational
bands of (superdeformed) nuclei.$^{31,32}$

To complete this work, it would be
interesting to study the
$U_{q}({\rm su}_2)$ CGc's in terms of
the $q$-deformed hypergeometric function
$_3F_2$ and to examine the contraction of
$_3F_2$ into a
$_2F_1$ operator-valued function when going from
the CGc to the Wigner operator. Also, to find $q$-boson
realizations for the Racah operator (the matrix elements
of which are $q$-deformed recoupling coefficients
or $6 - j$ symbols) is an
interesting problem. In addition, a treatment via symbolic
programming languages (like Maple or Mathematica)
of the formulas in this paper is presently under study. [A
programme is obtainable$^{34}$ for
calculating any $U_{q}({\rm su}_2)$
CGc.] We hope to return on these
matters in a future work.

\sb

Part of this work was presented (together with some
application to rotational spectroscopy) by one of
the authors (M.R.~K.)  to the Symposium ``Symmetries
in Science VIII''. The authors acknowledge
Prof.~B.~Gruber for giving them the opportunity
to present their results to this beautiful symposium.

\sb
\sd

\noindent {\bf 6. REFERENCES}

\sb

\item{1.} M. Kibler and G. Grenet,
% On the SU$_2$ unit tensor,
J. Math. Phys. 21:422 (1980).

\item{2.}
Yu.F. Smirnov and M.R. Kibler,
  Some aspects of $q$-boson calculus,
{\it in}: Symmetries in Science VI:
{}From the rotation group to quantum algebras,
B. Gruber, ed., Plenum Press, New York (1993), p.~691.

\item{3.}
M. Kibler, C. Campigotto and Yu.F. Smirnov,
   Recursion relations for Clebsch-Gordan coefficients
   of $U_q({\rm su}_2)$ and $U_q({\rm su}_{1,1})$,
{\it in}: Proceedings of the International
Workshop ``Symmetry Methods in Physics, in
Memory of Professor Ya.A. Smorodinsky'',
A.N. Sissakian, G.S. Pogosyan and S.I. Vinitsky, eds.,
J.I.N.R., Dubna, Russia (1994), p. 246.

\item{4.} E. Wigner, Gruppentheorie und ihre Anwendungen auf die
Quantenmechanik der Atomspektren, Vieweg \& Sohn, Braunschweig (1931).

\item{5.} B.L. van der Waerden, Die Gruppentheoretische
Methode in der Quanten\-mecha\-nik, Springer, Berlin (1932).

\item{6.} G. Racah,
%    Theory of complex spectra. II,
Phys. Rev. 62:438 (1942).

\item{7.} S.D. Majumdar,
%    The Clebsch-Gordan coefficients,
Prog. Theor. Phys. 20:798 (1958).

\item{8.} A.P.  Jucys and  A.A.  Bandzaitis,  Theory of angular
momentum in quantum mechanics, Mintis, Vilnius (1965).

\item{9.} L.C. Biedenharn and M. Tarlini,
%    On $q$-tensor operators for quantum groups,
Lett.~Math.~Phys.~20:271 (1990).

\item{10.} M. Nomura,
%    Recursion relations for the Clebsch-Gordan
%    coefficient of quantum group $SU_q(2)$,
J.~Phys.~Soc.~Jpn.~59:1954 (1990).

\item{11.} M. Nomura,
%    A Jordan-Schwinger representation of
%    quadratic relations for $SU_q(2)$ operators
%    and of the $q$-analog Wigner-Eckart theorem,
J.~Phys.~Soc.~Jpn.~59:2345 (1990).

\item{12.} V.A. Groza, I.I. Kachurik and A.U. Klimyk,
J. Math. Phys. 31:2769 (1990).

\item{13.} Yu.F. Smirnov, V.N. Tolsto\u \i~and Yu.I. Kharitonov,
%   Method of projection operators and the $q$ analog of the
%   quantum theory of angular momentum. Clebsch-Gordan
%   coefficients and irreducible tensor operators,
Sov. J. Nucl. Phys. 53:593 (1991).

\item{14.} C. Quesne,
    $q$-bosons and irreducible tensors for $q$-algebras,
{\it in}: Symmetries in Science VI:
{}From the rotation group to quantum algebras,
B. Gruber, ed., Plenum Press, New York (1993).

\item{15.} J. Shapiro,
%    Matrix representation of the angular
%    momentum projection operator,
J. Math. Phys. 6:1680 (1965).

\item{16.} J.-L. Calais,
%    Titre?
Int. J. Quantum Chem. 2:715 (1968).

\item{17.} A.J. Macfarlane,
%   On $q$-analogues of the quantum harmonic
%   oscillator and the quantum group $SU(2)_q$,
J.~Phys.~A 22:4581 (1989).

\item{18.} L.C. Biedenharn,
%   The quantum group $SU_q(2)$ and a
%   $q$-analogue of the boson operators,
J.~Phys.~A 22:L873 (1989).

\item{19.}
A. Sudbery,
%   Title?,
J. Phys. A: Math Gen. 23:L697 (1990).

\item{20.}
N. Reshetikhin,
%   Title?,
Lett. Math. Phys. 20:331 (1990).

\item{21.}
D.B. Fairlie and C.K. Zachos,
%   Title?,
Phys. Lett. B 256:43 (1991).

\item{22.}
A. Schirrmacher, J. Wess and B. Zumino,
%   Title?,
Z. Phys. C 49:317 (1991).

\item{23.}
S.T. Vokos,
%   Title?,
J. Math. Phys. 32:2979 (1991).

\item{24.}
R. Chakrabarti and R. Jagannathan,
%   Title?,
J. Phys. A: Math. Gen. 24:L711 (1991).

\item{25.} V.K. Dobrev,
%   Title?,
J. Math. Phys. 33:3419 (1992).

\item{26.}
Yu.F. Smirnov and R.F. Wehrhahn,
%   Title?,
J. Phys. A: Math. Gen. 25:5563 (1992).

\item{27.} M.R. Kibler,
  Introduction to quantum algebras,
{\it in}: Symmetry and Structural Properties of Condensed Matter,
W. Florek, D. Lipi\'nski and T. Lulek, eds.,
World Scientific, Singapore (1993), p. 445.

\item{28.}
S. Meljanac and M. Milekovi\'c,
%   Title?,
J. Phys. A: Math. Gen. 26:5177 (1993).

\item{29.}
C. Quesne,
%   Title?,
Phys. Lett. A 174:19 (1993).

\item{30.}
R. Chakrabarti and R. Jagannathan,
%   Title?,
J. Phys. A: Math. Gen. 27:2023 (1994).

\item{31.} R. Barbier, J. Meyer and M. Kibler,
% An $U_{qp}({\rm u}_2)$
% model for rotational bands of nuclei,
J. Phys. G: Nucl. Part. Phys. 17:L67 (1994).

\item{32.} R. Barbier, J. Meyer and M. Kibler,
A $qp$-rotor model for rotational
bands of superdeformed nuclei,
Preprint LYCEN 9437, IPNL (1994).

\item{33.}
Ya.A. Smorodinski{\u \i} and L.A. Shelepin,
% Clebsch-Gordan coefficients
% viewed from different sides,
Usp. Fiz. Nauk 15:1 (1972).

\item{34.} M. Kibler, G.-H. Lamot and Yu.F. Smirnov,
to be published.

\bye